\documentclass[aps,prl,twocolumn,showpacs,preprintnumbers,superscriptaddress]{revtex4}

\usepackage{graphicx}
\usepackage{epsf} 
\usepackage{dcolumn}
\usepackage{bm}
\usepackage{amssymb}
\usepackage{amsmath}
\usepackage{amsbsy}

\usepackage{graphicx,color,ifthen}
\usepackage{pstricks,pst-node,pst-text,pst-3d,color}

\newcommand{\Ge}{\Gamma_{ex}}
\newcommand{\Ga}{\Gamma_{ad}}

\begin{document}

\title{Non-linear ripple dynamics on amorphous surfaces patterned by ion-beam sputtering}

\author{Javier Mu\~noz-Garc\'{\i}a}
\affiliation{Departamento de Matem\'aticas and Grupo Interdisciplinar de
Sistemas Complejos (GISC), Universidad Carlos III de Madrid, Avenida de la
Universidad 30, E-28911 Legan\'es, Spain} 
\author{Mario Castro}
\affiliation{GISC and Grupo de Din\'amica No Lineal (DNL), Escuela T\'ec.\
Sup.\ de Ingenier{\'\i}a (ICAI), \\ Universidad Pontificia Comillas, E-28015
Madrid, Spain} 
\author{Rodolfo Cuerno}
\affiliation{Departamento de Matem\'aticas and Grupo Interdisciplinar de
Sistemas Complejos (GISC), Universidad Carlos III de Madrid, Avenida de la
Universidad 30, E-28911 Legan\'es, Spain} 

\date{\today}

\begin{abstract}
Erosion by ion-beam sputtering (IBS) of amorphous targets at off-normal
incidence frequently produces a (nanometric) rippled surface pattern, strongly
resembling macroscopic ripples on aeolian sand dunes. Suitable generalization
of continuum descriptions of the latter allows us to describe theoretically
for the first time the main nonlinear features of ripple dynamics by IBS,
namely, wavelength coarsening and non-uniform translation velocity, that agree
with similar results in experiments and discrete models. These properties are
seen to be the anisotropic counterparts of in-plane ordering and (interrupted)
pattern coarsening in IBS experiments on rotating substrates and at normal
incidence.
\end{abstract}

\maketitle

Ever since their earliest observation \cite{navez}, the production of ripples
on the surfaces of amorphous targets subject to ion-beam sputtering (IBS) at
intermediate energies, has been fascinating due to the similarities with
macroscopic ripples, like those produced underwater \cite{vortex_exp}, or on
the surface of aeolian sand dunes \cite{dunas}. Beyond the morphological
resemblance, IBS ripples share many other properties with e.g.\ aeolian
ripples, such as wavelength coarsening and pattern translation with time
\cite{carter,habenicht}. Remarkably, while typical wavelengths of the latter
are above 1 cm, the periodicity of IBS ripples is in the 100 nm range
\cite{valbusa}, these patterns having gained increased interest recently for
applications in Nanotechnology, ranging from optoelectronic to catalytic
\cite{applic}. IBS ripples are produced on a wide class of substrates, from
amorphous or amorphizable (silica, Si, GaAs, InP) to metallic targets (Cu, Au,
Ag) \cite{valbusa}. In view moreover of their implied loss of in-plane
symmetry, IBS ripples provide interesting instances of systems hosting a
competition between pattern forming and disordering mechanisms \cite{patt}.

A successful description of the main features of IBS ripples was provided by
Bradley and Harper (BH) \cite{BH}, based on Sigmund's linear cascade
approximation of sputtering processes in amorphous or polycrystalline targets
\cite{Sigmund}. The linear equation derived by BH describes satisfactorily
some properties of IBS ripples, such as their alignment with the ion beam as a
function of the incidence angle to target normal $\theta$ [wave vector
parallel (perpendicular) to the projection of the ion beam for $\theta <
\theta_c$ ($\theta > \theta_c$), for some threshold $\theta_c$].  Other
features, such as ripple stabilization or wavelength dependence with ion
energy or flux, required non-linear extensions of BH's approach
\cite{Makeev,park}, leading to an anisotropic generalization of the well known
Kuramoto-Sivashinsky (KS) equation \cite{patt,Makeev}. However, a notable
limitation of the anisotropic KS (AKS) equation is its inability to predict
ripples that coarsen with time, contradicting observations in many experiments
and/or discrete models of IBS (see \cite{habenicht,yewande} and Refs.\
therein).

In this Letter, we introduce a ``hydrodynamic'' model \cite{dunas,Aste,prl} for
IBS ripple production at off-normal incidence. Time scale separation between
microscopic processes (collision cascades, surface diffusion) and pattern
evolution allows us to derive an improved equation for the surface height.
The new non-linear terms appearing allow for ripple coarsening and pattern
translation with non-uniform velocity, as seen in experiments and discrete
models. Our theory has both the AKS equation \cite{Makeev} and the normal
incidence hydrodynamic theory \cite{prl,normal} as particular limits, and
enables analysis of the important case of rotating substrates \cite{frost}. In
addition, our model may be important also to the context of ripples on aeolian
sand dunes, where the standard 1D approximation requires validation, fully
anisotropic 2D models being scarce \cite{dunas}, as incidentally occurs in
many other contexts within Pattern Formation \cite{bar_nepomnyaschy2}.

During IBS of amorphous or semiconductor substrates, in which the ions
amorphize the subsurface layer, incident ions lose their energy through random
collision cascades in the bulk \cite{Sigmund}. Near-surface atoms receiving
enough energy and momentum to break their bonds are in principle sputtered
away, although they may join the current of surface adatoms
that are available to other relaxation mechanisms, such as surface diffusion, 
before incorporating back to the solid bulk. Within the so-called
``hydrodynamic'' approach to aeolian sand dunes \cite{dunas} and ion-sputtered 
surfaces \cite{Aste,prl}, we consider two coupled fields, namely, the thickness
of the mobile surface adatoms layer, $R(\mathbf{x},t)$, (related with the 
density of mobile adatoms through the atomic volume) and the height of the 
bombarded surface above a reference plane, $h(\mathbf{x},t)$. 
Their time evolution is provided by 
\begin{align}
        \partial_t R &= (1-\phi) \Gamma_{ex} - \Gamma_{ad} + D \nabla^2 R \label{eq.R},\\
        \partial_t h &= -\Gamma_{ex}+\Gamma_{ad} \label{eq.h} ,
\end{align}
where $\Ge$ and $\Ga$ are, respectively, rates of atom excavation from and
addition to the immobile bulk, $(1-\phi)=\bar{\phi}$ measures the fraction of
eroded atoms that become mobile, and the third term in Eq.\ (\ref{eq.R})
describes motion of mobile atoms along the surface as due to isotropic thermal
diffusion ($D$ being a constant for amorphous materials). Even if all eroded
atoms are sputtered away ($\phi=1$), we assume a non-zero average fraction of
mobile atoms, $R_{eq}$.

Under the assumption that nucleation events are more likely in surface        
protrusions, in analogy to the Gibbs-Thompson relation we have          
\begin{equation}\label{Ga}
\Gamma_{ad}=\gamma_0\left[R(1+\gamma_{2x}\partial^2_xh+\gamma_{2y}\partial^2_yh)-R_{eq}
\right],
\end{equation}
where $\gamma_0$ is the mean nucleation rate for a flat surface (on the
$xy$ plane) and $\gamma_{2x}$, $\gamma_{2y} \geq 0$ describe variation of the
nucleation rate with surface curvatures. Note that, in (\ref{Ga}), the full
thickness of the mobile adatoms layer is affected by the shape of the surface.

The rate at which material is sputtered from the bulk depends on the angle of
incidence, ion and substrate species, ion flux, energy, and many other
experimental parameters. If the beam direction is in the $xz$ plane, we have,
following \cite{Makeev,prl}
\begin{multline}\label{Ge}
    \Gamma_{ex} = \alpha_0 \big[ 1 + \alpha_{1x} \partial_x h+ \alpha_{2x} \partial^2_x h
    +\alpha_{2y} \partial^2_y h \\
    +\alpha_{3x}(\partial_x h)^2+\alpha_{3y} (\partial_y h)^2-(\partial_x h) (\alpha_{4x}
    \partial^2_x h + \alpha_{4y} \partial^2_y h)\big],
\end{multline}
where parameters reflect the dependence of $\Gamma_{ex}$ on the {\em local} 
shape of the surface \cite{detalles}, as described by more microscopic 
derivations such as BH or generalizations thereof \cite{Makeev,Feix}. 
Analogous, but not equal, geometrical couplings to the driving
occur in aeolian sand dunes \cite{dunas}, or in {\em growth} onto amorphous 
substrates \cite{Raible}. Note the loss of reflection symmetry in the $x$ 
direction, but not in the $y$ direction. For a planar surface, atoms are 
sputtered from the bulk in a typical time of order $\alpha_0^{-1}$.

The main difference between former models \cite{BH,Makeev,park} and our
present model, Eqs.\ (\ref{eq.R})-(\ref{Ge}), is that, in the latter, eroded
material is allowed to redeposit locally, and there is an implicit viscous flow
\cite{viscousflow} in the amorphized layer through the evolution of $R$.
These {\em additional mechanisms} are seen below to induce richer
pattern dynamics than in \cite{BH,Makeev,park}.

The linearized Eqs.\ (\ref{eq.R})-(\ref{Ge}) have solutions ${R}^l = \hat{R^l}
e^{i\mathbf{k\cdot x} +\omega_{\mathbf{k}} t}$, $h^l=\hat{h^l}
e^{i\mathbf{k\cdot x}+\omega_{\mathbf{k}} t}$, with a dispersion relation,
$\omega_{\mathbf{k}}$, given in the long wavelength limit by~\cite{detalles}
\begin{eqnarray} 
   \ \mathcal{R}e (\omega_{\mathbf{k}})&=& \epsilon \phi \gamma_0(\alpha_{2x}k_x^2+
    \alpha_{2y}k_y^2)  -\epsilon^2 \phi \bar{\phi}\gamma_0 \alpha_{1x}^2 k_x^2  \nonumber \\
     \lefteqn{-R_{eq}D(k_x^2+k_y^2)(\gamma_{2x}k_x^2+
    \gamma_{2y}k_y^2)+\mathcal{O}(\epsilon k^4) , }   \qquad \label{rel.disp.}  \\ 
\mathcal{I}m(\omega_{\mathbf{k}}) &=&-\epsilon \phi \gamma_0 \alpha_{1x} k_x+ 
\mathcal{O}(k^3), \label{im.rel.disp.} 
\end{eqnarray}
where $\epsilon\equiv\alpha_0 / \gamma_0$ is a dimensionless parameter; the
erosion rate being much smaller than the nucleation rate, $\epsilon \ll 1$ for
typical experiments. Eq.\ (\ref{im.rel.disp.}) is a simple consequence of the
asymmetry in the $x$ direction, induced by the incoming flux.
                                                       
The surface morphology is dominated by the periodic pattern with wave vector
$\mathbf{k}^{max}$ making Eq.\ \eqref{rel.disp.} a positive maximum.  
It can be shown \cite{detalles} that $\mathbf{k}^{max}$ is oriented along the
$\mathbf{\hat{x}}$ or $\mathbf{\hat{y}}$ directions, as observed
experimentally \cite{valbusa}. Close to the instability threshold, before
nonlinear terms are no longer negligible, one has $k_{x,y}^{max}\sim
\epsilon^{1/2}$. Substituting this into Eqs.\ \eqref{rel.disp.} and
\eqref{im.rel.disp.} provides us with estimations of the typical time and
length scales of the pattern, that we employ to rescale $X=\epsilon^{1/2}x$,
$Y=\epsilon^{1/2}y$, $T_2=\epsilon^2 t$ and $T_1=\epsilon^{3/2}t$, and perform
a multiple scale expansion of Eqs.\ \eqref{eq.R}, \eqref{eq.h}, in which $R$
can be adiabatically eliminated. To lowest non-linear order, we get
\cite{detalles, nota_no_redep}
\begin{align}
\partial_t h &= \gamma  \partial_x h + \sum_{i=x,y} \Bigl\{
-\nu_i \, \partial^2_i h + \lambda^{(1)}_{i} \, (\partial_i h)^2 + \Omega_ i \partial_i^2 \partial_x h \nonumber \\
+ \xi_i &(\partial_x h) (\partial^2_i h)\Bigr\}+ \sum_{i,j=x,y}\Bigl\{ -{\cal K}_{ij} \partial^2_i \partial^2_j h + \lambda^{(2)}_{ij}
\partial^2_{i} (\partial_j h)^2 \Bigr \} ;\nonumber \\
	\gamma&=-\epsilon \phi \gamma_0 \alpha_{1x}, \quad
	\nu_x= \epsilon \phi \gamma_0 \alpha_{2x} - \epsilon^2 \bar{\phi} \phi \gamma_0 
	\alpha_{1x}^2, \nonumber \\
	\nu_y &= \epsilon \phi \gamma_0 \alpha_{2y},\quad
	\lambda_{i}^{(1)}=-\epsilon \phi \gamma_0 \alpha_{3i}, \nonumber \\
	\Omega_i &= \epsilon (\bar{\phi}D 
	- \phi R_{eq} \gamma_0 \gamma_{2i}) \alpha_{1x},  \quad  \xi_i =
	\epsilon \phi \gamma_0 \alpha_{4i}, \nonumber \\
	{\cal K}_{ij}&= D R_{eq} \gamma_{2i} + \epsilon \big[ D \bar{\phi} 
	(\gamma_{2i} -\alpha_{2j}) +\phi \gamma_0 R_{eq}\gamma_{2i} 
	\alpha_{2j} \big],\nonumber \\
	\lambda^{(2)}_{ij}&=\epsilon  \big[ \bar{\phi} D - \phi R_{eq} \gamma_0 
	\gamma_{2i} \big] \alpha_{3j} . \label{eq.ero} 
\end{align}
Eq.\ (\ref{eq.ero}) generalizes the AKS type equations obtained
\cite{BH,Makeev,park} within BH approach to IBS. While sharing the same
reflection properties in the $x$ and $y$ directions and most of the terms on
the rhs, both equations differ crucially by the presence here of the
$\lambda^{(2)}_{ij}$ nonlinearities. Moreover, in the absence of redeposition
($\phi=1$), $\lambda^{(1)}_i$ and $\lambda^{(2)}_{ij}$ have the same signs, 
making Eq.\ \eqref{eq.ero} nonlinearly unstable, 
as in the BH case \cite{prl,Kim,comments}.
Note, the linear dispersion relation of (\ref{eq.ero}) matches Eqs.\
\eqref{rel.disp.} and \eqref{im.rel.disp.} above. Under normal incidence,
parameters are isotropic and $\alpha_{1x}=\alpha_{4x}=\alpha_{4y}=0$, Eq.\
\eqref{eq.ero} reducing to that obtained in \cite{prl, Kim}, and in
\cite{Raible}. 
As a final remark,
let us quote the form of Eq.\ \eqref{eq.ero} for {\em sample rotation} around
the $z$ axis during bombardment (see e.g.\ \cite{Bradley,frost}). Dynamics of
$h$ are given by a different isotropic limit of Eq.\ (\ref{eq.ero}), namely,
\begin{align}
 \partial_t h &= - \nu_{\rm r} \nabla^2 h - 
 {\cal K}_{\rm r} \nabla^4 h +\lambda^{(1)}_{\rm r} (\nabla h)^2 + \lambda^{(2)}_{\rm r} \nabla^2(\nabla h)^2    \nonumber\\
&+\lambda^{(3)}_{\rm r} \, \nabla \cdot \big[ (\nabla^2 h) \nabla h \big] \big] ; \label{eq.rotante} \\
	\nu_{\rm r}&= (\nu_x+\nu_y)/2, \quad
	\lambda^{(1)}_{\rm r}=(\lambda^{(1)}_x+\lambda^{(1)}_y)/2,\nonumber\\
		\lambda^{(2)}_{\rm r}&=\frac{1}{4} \sum_{i,j=x,y} \lambda^{(2)}_{i,j},\quad
	\lambda^{(3)}_{\rm r}= \frac{1}{2} \sum_{i,j=x,y} \lambda^{(2)}_{i,j} 
	\delta_{i,j} -\lambda^{(2)}_{\rm r},  \nonumber \\
	\cal K_{\rm r}&=(3{\cal K}_{x,x}+3{\cal K}_{y,y}+{\cal K}_{x,y}+{\cal K}_{y,x})/8,\nonumber
\end{align}
with parameters (susbscript ``r'' denotes ``rotating'').

\begin{figure}[!t]
\begin{minipage}{0.4\linewidth}
\includegraphics[width=\textwidth]{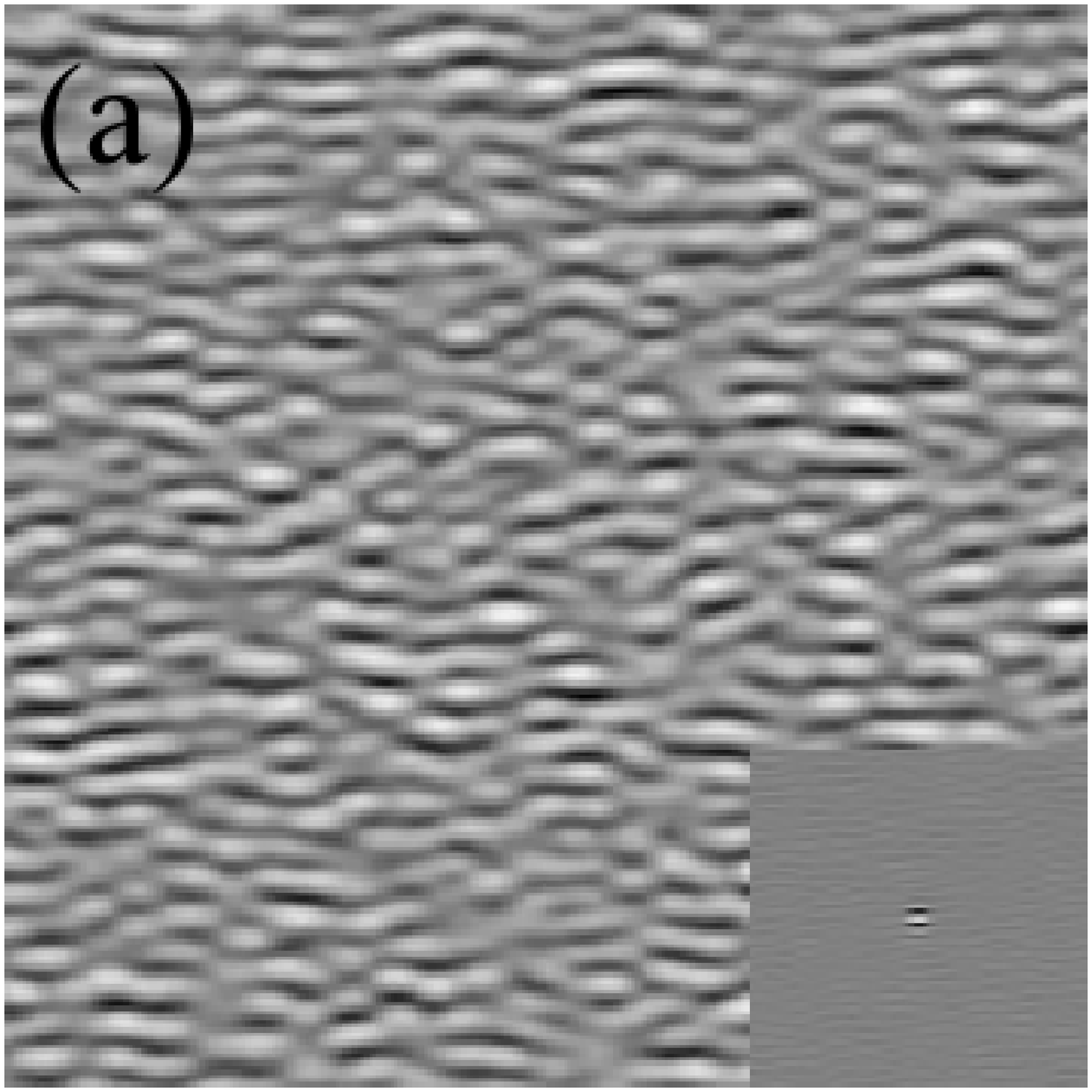}
\end{minipage}
\hspace{0.15cm}
\begin{minipage}{0.4\linewidth}
\includegraphics[width=\textwidth]{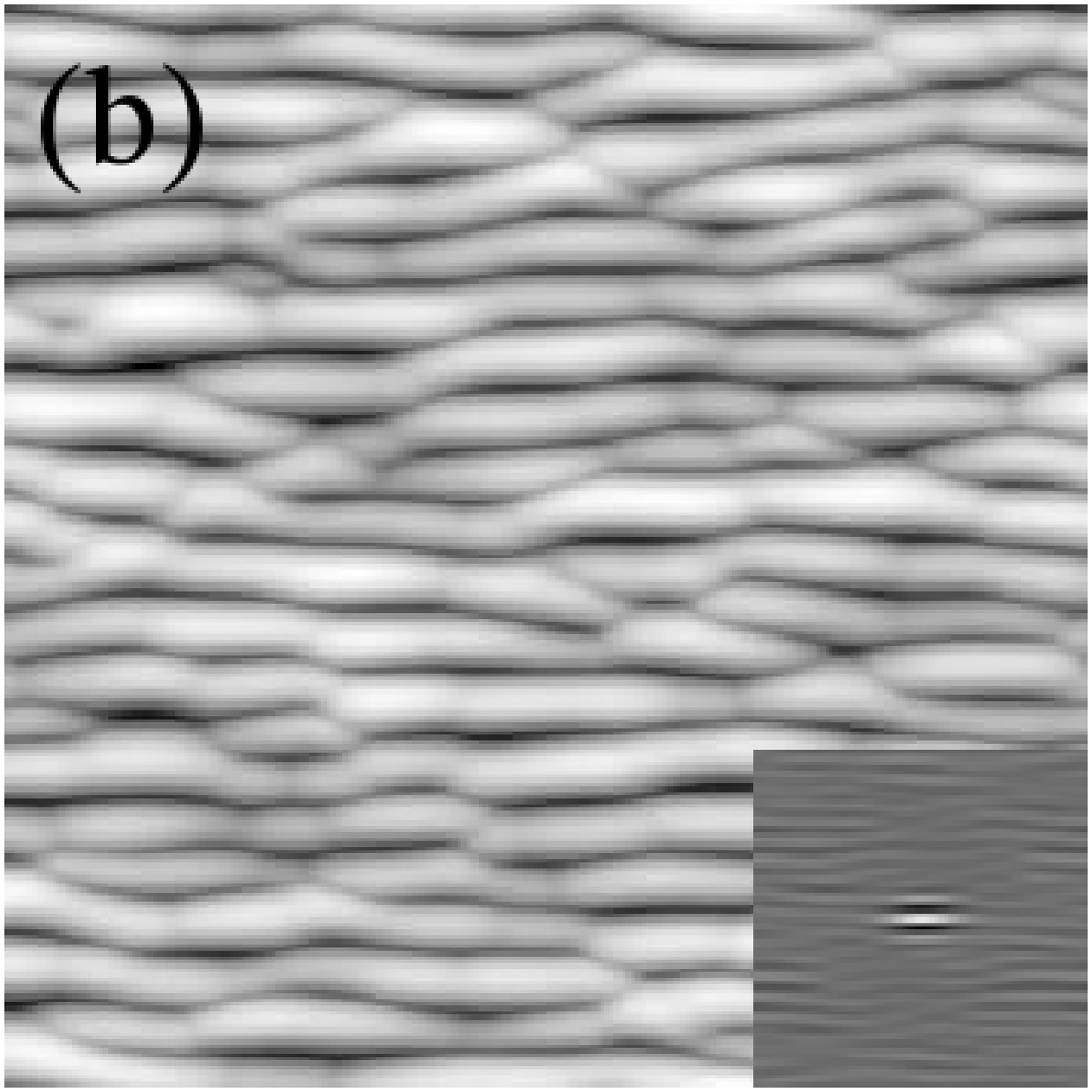}
\end{minipage}

\vspace{0.1cm}

\begin{minipage}{0.4\linewidth}
\includegraphics[width=\textwidth]{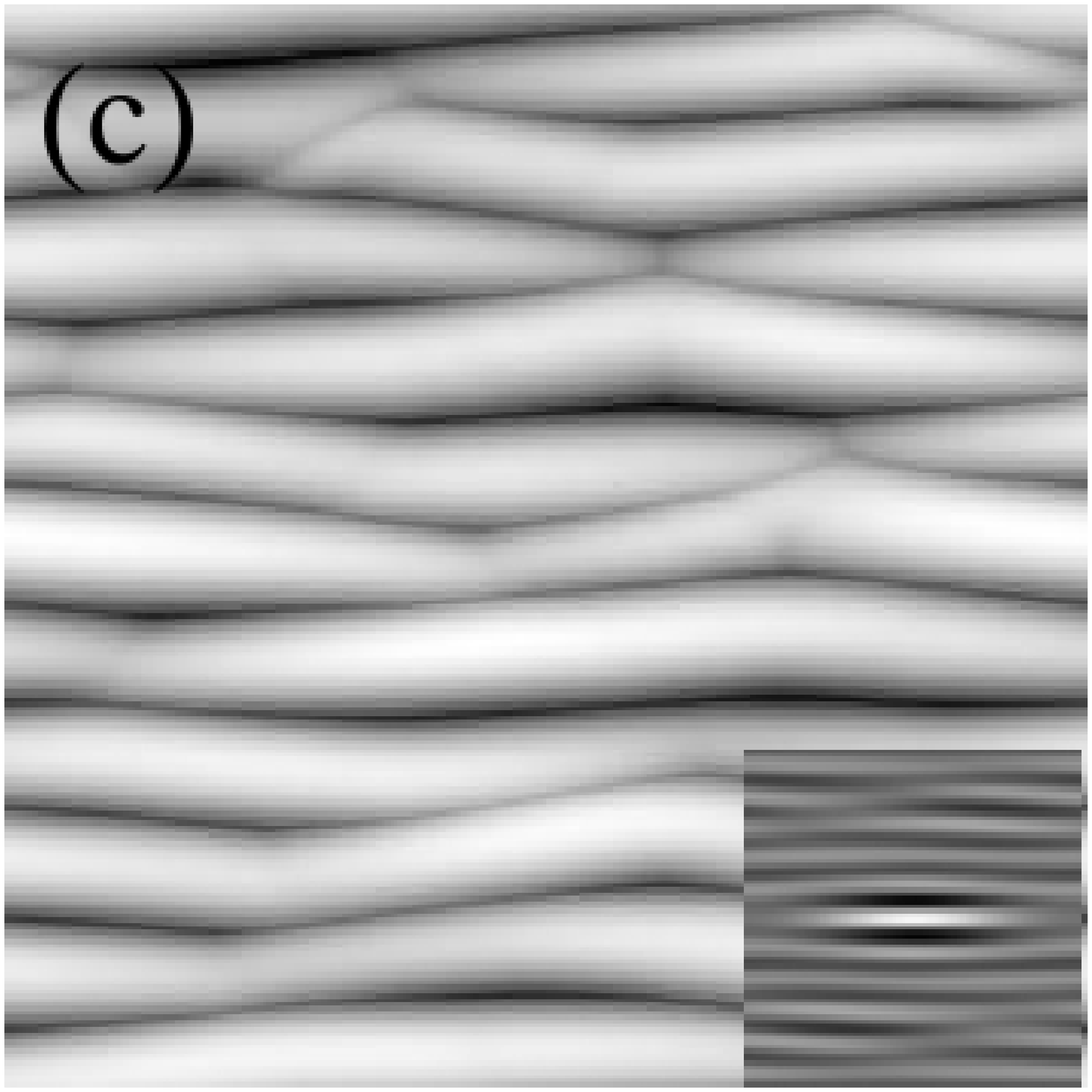}
\end{minipage}
\hspace{0.015cm}
\begin{minipage}{0.425\linewidth}
\includegraphics[width=\textwidth]{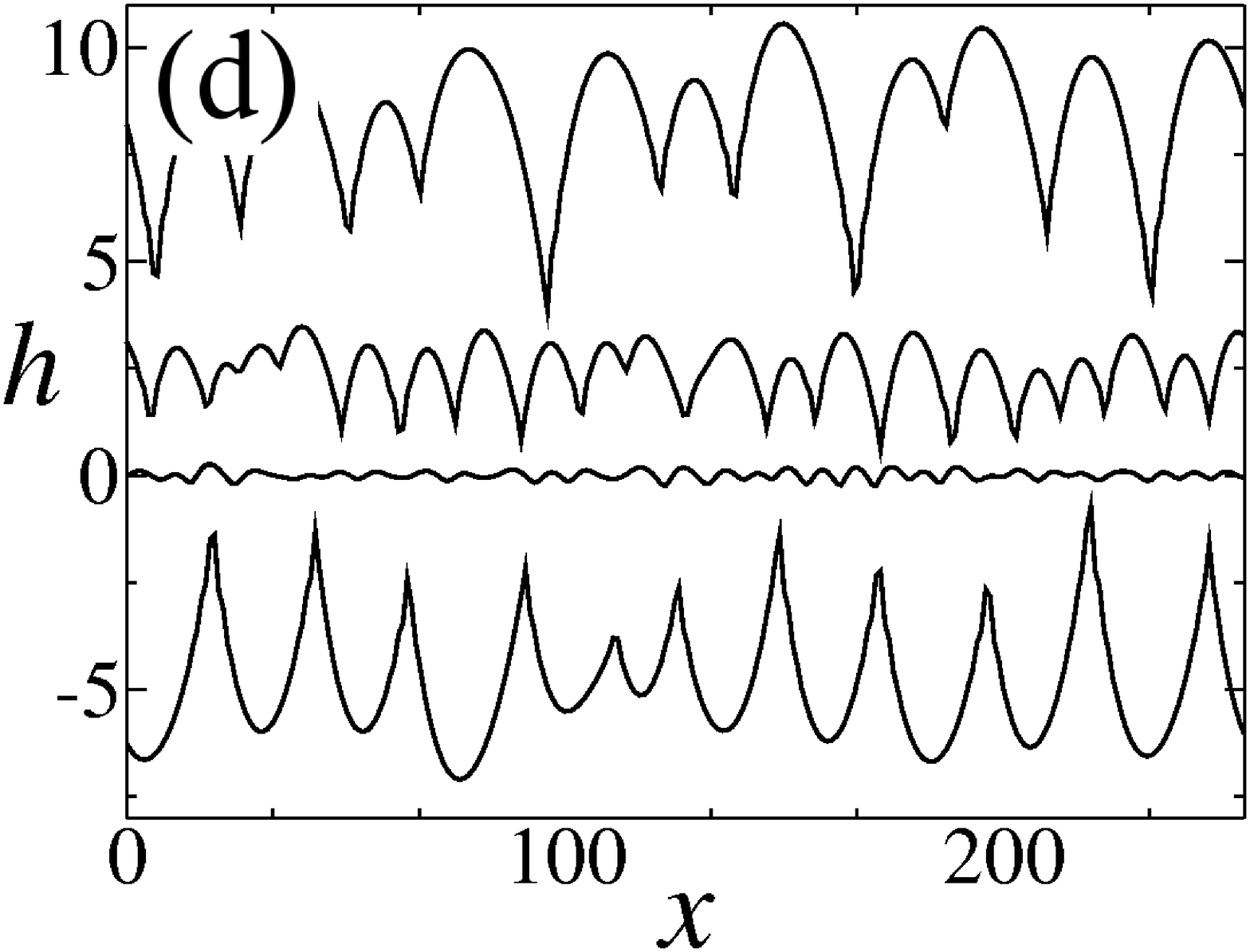}
\vspace{-0.55cm}
\hspace{0.135cm}
\end{minipage}
\caption{Numerical integration of Eq.\ \eqref{eq.ero} using $\nu_x=1$,
$\nu_y=0.1$, $\Omega_x=1$, $\Omega_y=0.5$, $\xi_i=0.1$, $\lambda_x^{(1)}=0.1$,
$\lambda_y^{(1)}=5$, $\lambda^{(2)}_{i,j}=-5$, ${\cal K}_{i,j}=1$. Top views
for $t= 10, 106, 953$ $(a)$, $(b)$, and $(c)$, respectively. Insets are height
autocorrelations. All units are arbitrary.
$(d)$: Top to bottom, side cuts of $(c)$, $(b)$, $(a)$, and
grooves obtained for sign-changed $\lambda^{(1)}_{x,y}$ and
$\lambda^{(2)}_{i,j}$. Curves have an artificial offset.}
\label{fig1}
\end{figure}

\begin{figure}[!t]
\begin{minipage}{0.45\linewidth}
\includegraphics[width=\textwidth]{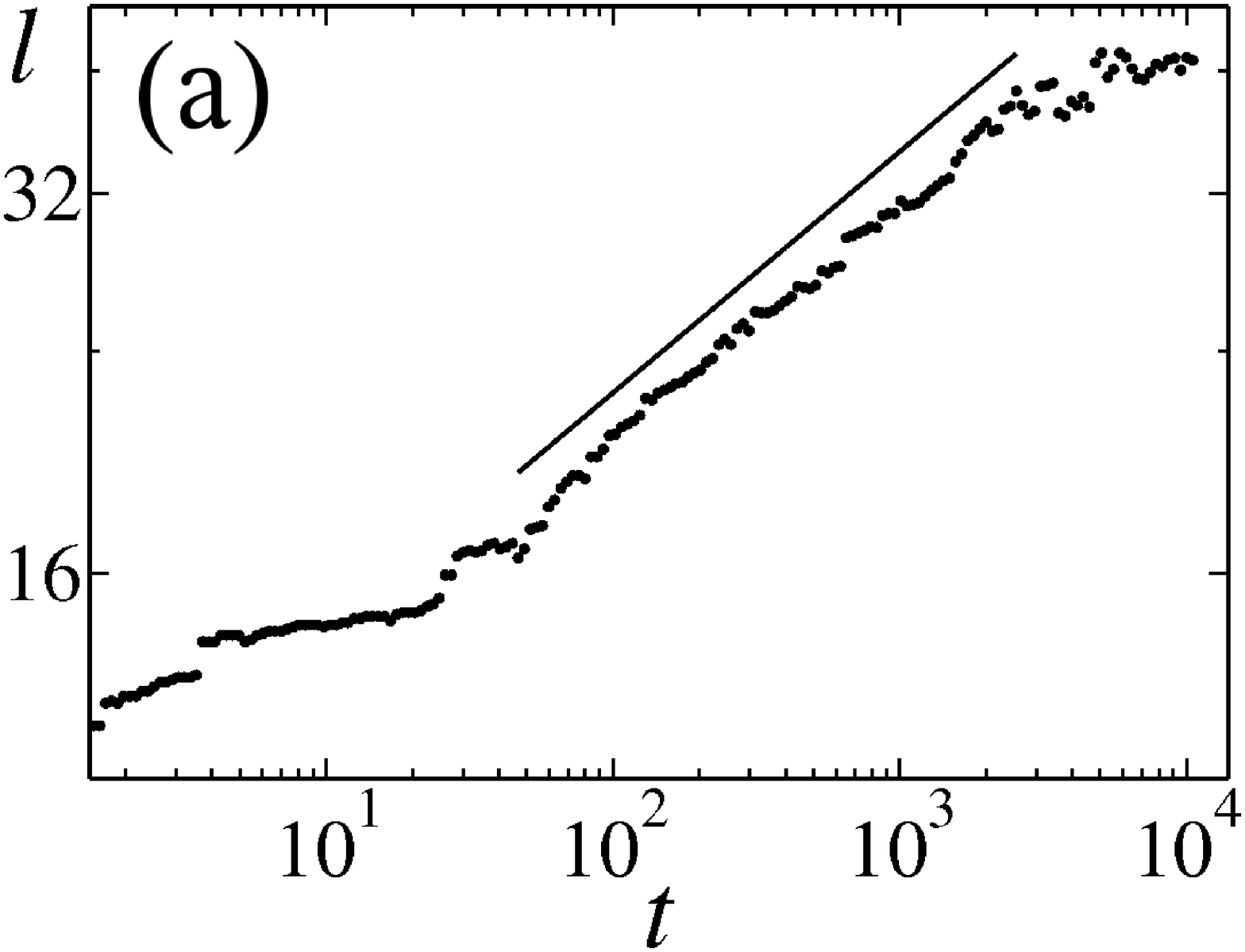}
\end{minipage}
\hspace{0.15cm}
\begin{minipage}{0.45\linewidth}
\includegraphics[width=\textwidth]{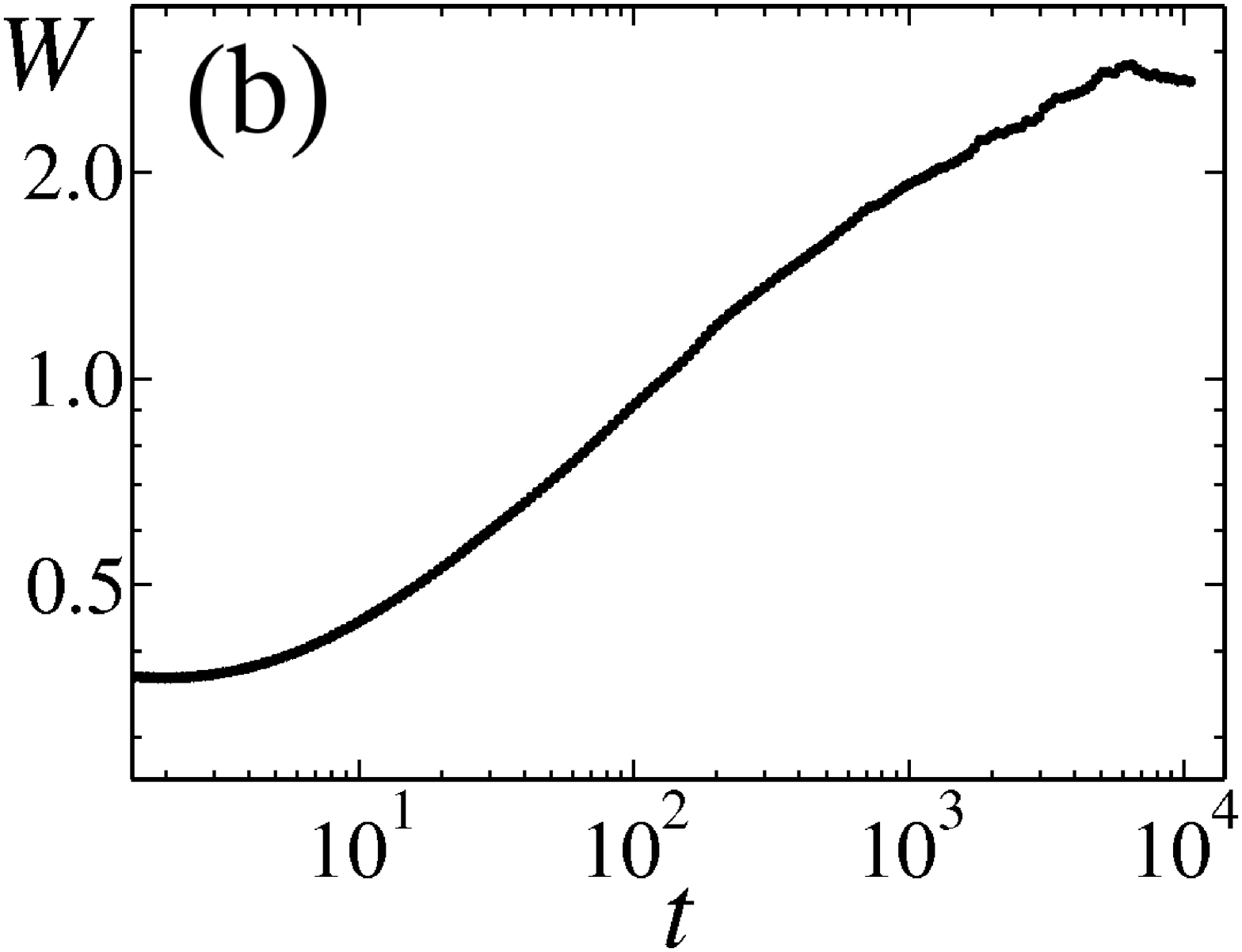}
\end{minipage}
\caption{Ripple wavelength $\ell(t)$ and surface roughness $W(t)$ for 
parameters as in Fig.\ \ref{fig1}. The straight line represents 
$\ell \sim t^{0.19}$. All units are arbitrary.}
\label{fig2}
\end{figure}

\begin{figure}[!thbp]
\begin{minipage}{0.4\linewidth}
\includegraphics[width=\textwidth]{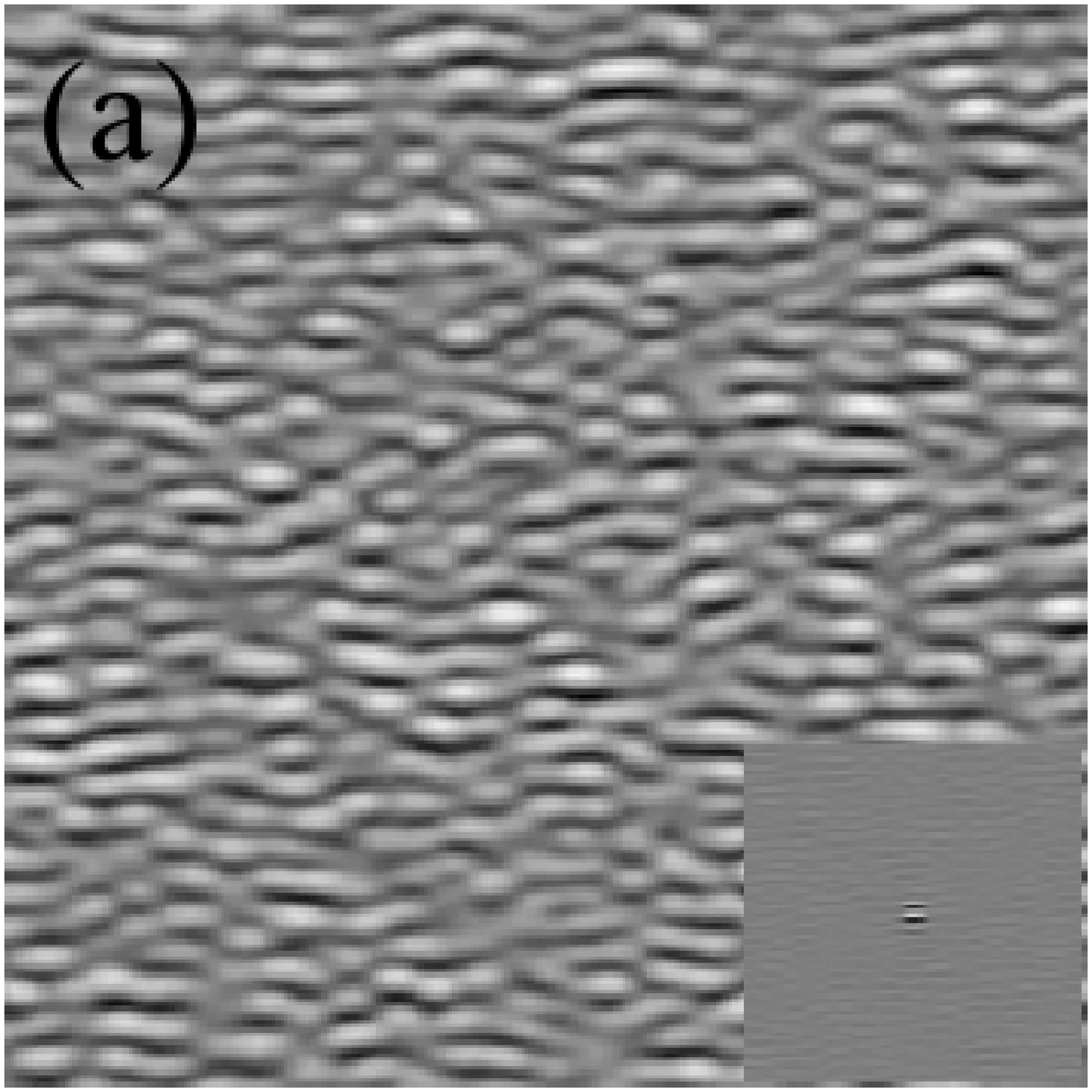}
\end{minipage}
\hspace{0.15cm}
\begin{minipage}{0.4\linewidth}
\includegraphics[width=\textwidth]{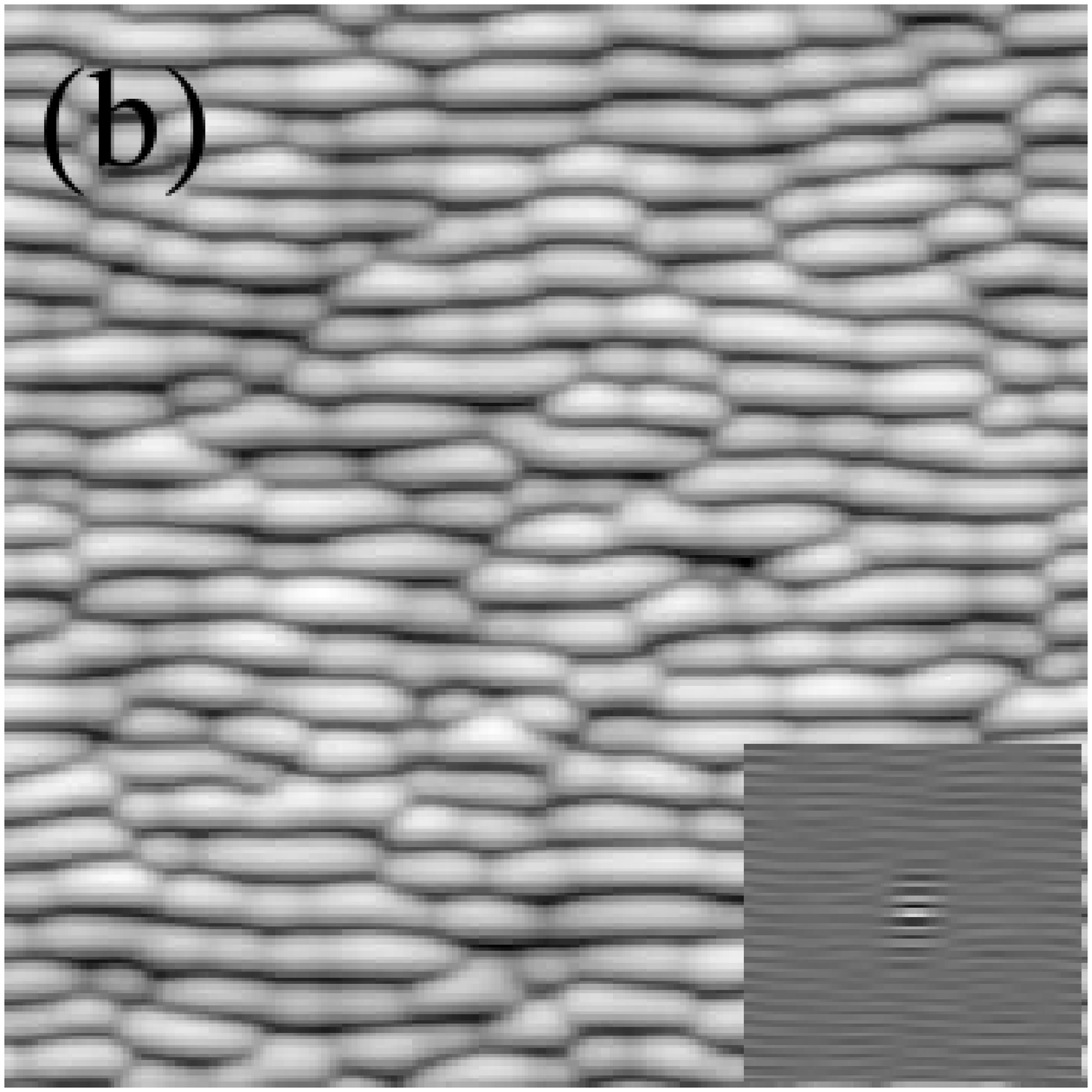}
\end{minipage}

\vspace{0.1cm}

\begin{minipage}{0.4\linewidth}
\includegraphics[width=\textwidth]{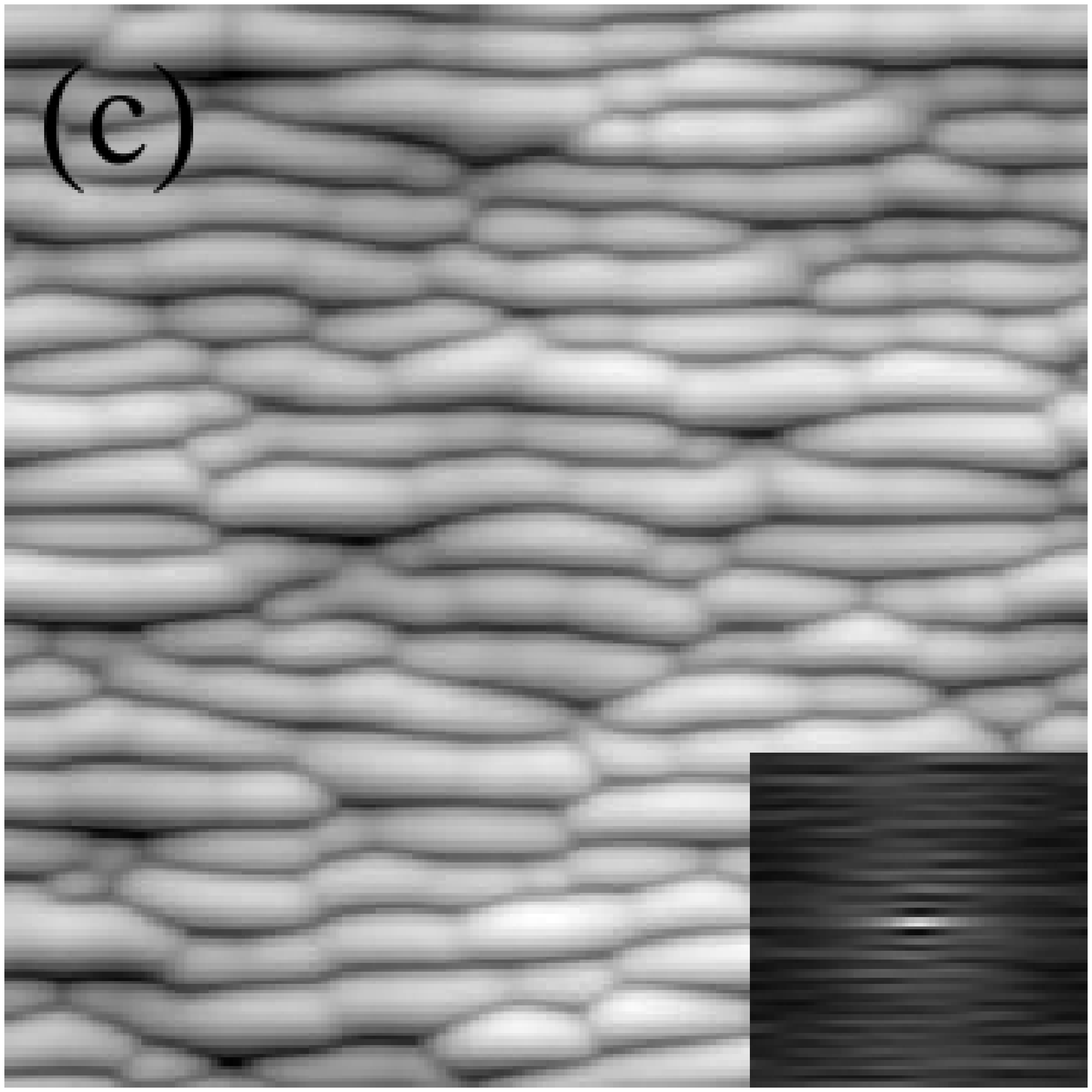}
\end{minipage}
\hspace{0.015cm}
\begin{minipage}{0.425\linewidth}
\includegraphics[width=\textwidth]{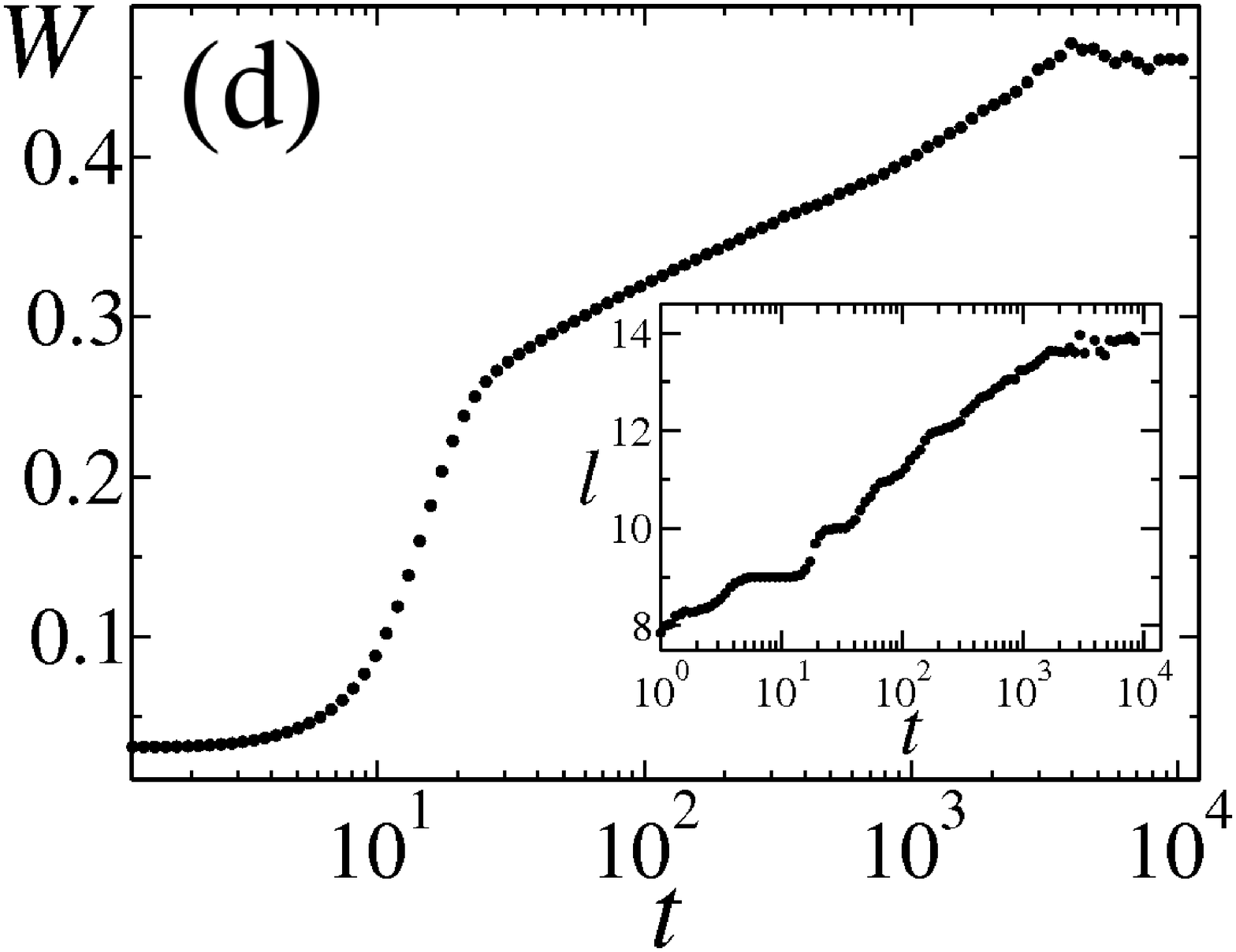}
\vspace{-0.55cm}
\hspace{0.135cm}
\end{minipage}
\caption{Numerical integration of Eq.\ \eqref{eq.ero} for parameters as in
Fig.\ \ref{fig1}, except for $\lambda_x^{(1)}=1$. Top views for $t= 10, 106, 
953$ $(a)$, $(b)$, and $(c)$, respectively. $(d)$: Surface roughness 
$W(t)$ for the same simulation. Inset: $\ell(t)$ for the same system, showing
marginal coarsening. All units are arbitrary.}
\label{fig3}
\end{figure}

To the best of our knowledge, Eq.\ (\ref{eq.ero}) is new, and indeed has a
rich parameter space. Numerical integration (not shown) within linear regime
retrieves all features of the ripple structure as predicted by the BH theory,
i.e., dependence of the ripple wavelength with linear terms, and ripple
orientation as a function of $\theta$. Entering the nonlinear regime, and as
occurs in the AKS equation and its generalizations \cite{Makeev,park},
nonlinearities $\lambda^{(1)}_i$ lead to saturation of the pattern with
constant wavelength and amplitude. In absence of these terms, the ripple
wavelength grows indefinitely as $\ell(t) \sim t^{n}$ with $n=1/2$
\cite{Raible_b, nota_exponente} until a single ripple remains in a finite
simulation domain. As an anisotropic generalization of the ordering process
observed for normal incidence \cite{prl, normal}, pattern coarsening requires
the presence of $\lambda^{(2)}_{i,j}$, whose magnitude and mathematically
correct sign \cite{comments} are due to describing redeposition by means of
the additional field $R$. When the values of these coefficients increase
relative to $\lambda^{(1)}_i$, coarsening stops later, and the amplitude and
wavelength of the pattern also increase. The coarsening exponent $n$ will take
an {\em effective} value that will be larger the later coarsening stops, and 
may depend on simulation parameter values.
For instance, we show in Fig.\ \ref{fig1} snapshots of a numerical integration 
of Eq.\ (\ref{eq.ero}) for relatively large $\lambda^{(2)}_{i,j}$. The apparent 
coarsening is quantified in the plot of $\ell(t)$ shown in 
Fig.\ \ref{fig2}$(a)$, compatible (after transient effects analogous to those 
in \cite{Raible_b}) with $n=0.19$.  Note the saturation of ripple
wavelength at long times, together with saturation in amplitude, as shown in
the plot of the surface roughness (rms width) $W(t)$ in Fig.\
\ref{fig2}$(b)$. These results are similar to those obtained experimentally
for IBS of Si in \cite{carter}. For a different experiment on Si, precise
measurements of the coarsening exponent \cite{habenicht} yield $n=0.50(4)$, no
saturation having been observed in this system, as we would expect. Here, the
dispersion velocity of the pattern was also measured, finding that it decays
with ripple wavelength or, equivalently, with time. We have also observed the
same trend in the dispersion velocity of the ripples shown in Fig.\
\ref{fig1}.  Experiments also exist, e.g.\ for IBS of Si, in which ripple
coarsening is absent or residual \cite{erlebacher}, that would correspond to
smaller $\lambda^{(2)}_{i,j}$ values in \eqref{eq.ero}, see e.g.\ an example
in Fig.\ \ref{fig3}, where $\ell(t) \sim \log t$ approximately. 
Ripple coarsening has been also observed in a Monte Carlo
model of IBS \cite{yewande}, in which rules implement Sigmund's theory. To our
purpose, the main conclusion of this study is the correlation between an
increasing $\ell(t)$ and non-uniform dispersion velocity, and on the
parameter-dependent values of the coarsening exponent. Let us note that, on
the experimental side, values of $n$ show a large scatter in the literature
(see Refs.\ in \cite{valbusa,yewande}).
Additional non-linear effects can be described by Eq. \eqref{eq.ero}.  A {\em
first} one is production of grooves (as opposed to ripples), due to the loss
of up-down symmetry induced by quadratic nonlinearities. Indeed, by changing
the sign of $\lambda_i^{(1)}$, grooves replace ripples, see Fig.\
\ref{fig1}$(d)$; this calls for systematic experimental exploration.  A {\em
second} effect is related with cancellation modes (CM), known in the AKS
equation \cite{rost_krug,park} and other models of IBS \cite{Aste, prl, Kim}.
These are linearly unstable modes for which nonlinearities cancel one
another exactly. While for equations like the one considered in \cite{prl,
Raible} CM affect well-posedness, anisotropic systems
\cite{rost_krug,park} may remain better defined in the presence of CM. In our
case, if reflection symmetry breaking terms are neglected in Eq.\
\eqref{eq.ero} and BH parameters are used, the same CM occur as in the AKS
equation. Additional CM ensue between the $\lambda_i^{(1)}$ and
$\lambda^{(2)}_{i,j}$ terms for appropriate relative signs 
\cite{Raible,comments}. We have verified
numerically that the {\em full} Eq.\ \eqref{eq.ero} breaks down for the latter
CM, but can support AKS-type CM. That these solutions are physically
realizable or are artifacts of the small slope approximation made,
remains to be assessed. Useful information on this
issue might come from field experiments in aeolian sand dunes.

In closing, we mention IBS of metals as an immediate experimental domain to
which the above results may be relevant \cite{valbusa}, albeit differing in
the degree of universality. There, however, the correct extension of BH theory
is not yet clear, nor is its importance relative to anisotropic surface
diffusion. We have taken preliminar steps in this context \cite{Feix}, and
expect to make progress in this direction soon.

\begin{acknowledgments}
We thank L. V\'azquez and R.\ Gago for discussions.  This work has been
partially supported by MECD (Spain), through Grants Nos.\ BFM2003-07749-C05,
-01 and -05, and an FPU fellowship (J.M.-G.).

\end{acknowledgments}



\begin{thebibliography}{99}

\bibitem{navez} M. Navez, C. Sella, and D. Chaperot, C. R. Acad. Sci. Paris
{\bf 254}, 240 (1962).

\bibitem{vortex_exp} A.\ Stegner and J.\ E.\ Wesfreid, Phys. Rev. E {\bf 60},
R3487 (1999).

\bibitem{dunas} O. Terzidis, P. Claudin and J.-P. Bouchaud, Eur. Phys. J. B
{\bf 5}, 245 (1998); A. Valance and F. Rioual, Eur. Phys. J. B {\bf 10}, 543
(1999); Z. Csah\'ok {\em et al.}, Eur. Phys. J. E {\bf 3}, 71 (2000).

\bibitem{carter} G. Carter and V. Vishnyakov, Phys. Rev. B {\bf 54}, 17647
(1996).

\bibitem{habenicht} S. Habenicht {\em et al.}, 
A. D. Wieck, Phys. Rev. B {\bf 65}, 115327 (2002).

\bibitem{valbusa} U. Valbusa, C. Boragno, and F. Buatier de Mongeot,
J. Phys. Condens. Matter {\bf 14}, 8153 (2002).

\bibitem{applic} M. V. R. Murty, Surf. Sci. {\bf 500}, 523 (2002); O. Azzaroni
{\em et al.}, Adv. Mater. {\bf 16}, 405 (2004).

\bibitem{patt} M. C. Cross and P. C. Hohenberg, Rev. Mod. Phys. {\bf 65}, 851
(1993).

\bibitem{BH} R. M. Bradley and J. M. Harper, J. Vac. Sci.  Technol. A {\bf 6},
2390 (1988).

\bibitem{Sigmund} P. Sigmund, Phys. Rev. {\bf 184}, 383 (1969);
J. Mat. Sci. {\bf 8}, 1545 (1973).

\bibitem{Makeev} R. Cuerno and A.-L. Barab\'asi, Phys. Rev. Lett. {\bf 74},
4746 (1995); M. Makeev, R. Cuerno, and A.-L. Bar\'abasi,
Nucl. Instrum. Methods Phys. Res., Sect. B {\bf 197}, 185 (2002).

\bibitem{park} S. Park {\em et al.} 
Phys. Rev. Lett. {\bf 83}, 3486 (1999).

\bibitem{yewande} E. Yewande, A. K. Hartmann, and R. Kree, Phys. Rev. B {\bf
71}, 195405 (2005).

\bibitem{Aste} T. Aste and U. Valbusa, Physica A {\bf 332}, 548 (2004); New
J. Phys. {\bf 7}, 122 (2005).

\bibitem{prl} M. Castro {\em et al.}, Phys. Rev. Lett. {\bf 94}, 016102
(2005).

\bibitem{normal} S. Facsko {\em et al.}, Science {\bf 285}, 1551 (1999);
R. Gago {\em et al.}, Appl. Phys. Lett. {\bf 78}, 3316
(2001). 

\bibitem{frost} F. Frost, A. Schindler, and F. Bigl, Phys. Rev. Lett.  {\bf
85}, 4116 (2000).

\bibitem{bar_nepomnyaschy2} D. E. Bar and A. A. Nepomnyashchy, Physica D {\bf
132}, 411 (1999).

\bibitem{detalles} J. Mu\~noz-Garc\'{\i}a, M. Castro, and R. Cuerno, {\em in
preparation}.

\bibitem{Feix} M. Feix {\em et al.}, Phys. Rev. B {\bf 71}, 125407 (2005).

\bibitem{Raible} M. Raible {\em et al.},
Europhys. Lett. {\bf 50}, 61 (2000); M. Raible, S. J. Linz, and P.
H\"{a}nggi, Phys. Rev. E {\bf 64}, 031506 (2001).

\bibitem{viscousflow} C.\ C. Umbach {\em et al.}, Phys. Rev. Lett. {\bf 87},
246104 (2001).

\bibitem{nota_no_redep} In principle, Eq.\ \eqref{eq.ero} does not cover the
conserved $\phi=0$ limit (irrelevant to IBS), where a different expansion must
be used, as seen from Eqs.\ \eqref{rel.disp.}, \eqref{im.rel.disp.}.

\bibitem{Kim} T. C. Kim {\em et al.}, Phys. Rev. Lett. {\bf 92}, 246104
(2004).

\bibitem{comments} M. Castro and R. Cuerno, Phys. Rev. Lett. {\bf 94}, 139601
(2005); T. C. Kim {\em et al.}, {\em ibid}. {\bf 94}, 139602 (2005).

\bibitem{Bradley} R. M. Bradley, Phys. Rev. E {\bf 54}, 6149 (1996).

\bibitem{Raible_b} M. Raible, S. J. Linz, and P. H\"{a}nggi, Phys. Rev. E {\bf
62}, 1691 (2000).

\bibitem{nota_exponente} This is in absence of advective terms. Otherwise,
even larger exponent values are possible \cite{detalles}.

\bibitem{erlebacher} J. Erlebacher {\em et al.}, Phys. Rev. Lett. {\bf 82},
2330 (1999).

\bibitem{rost_krug} M. Rost and J. Krug, Phys. Rev. Lett. {\bf 75}, 3894
(1995).

\end{thebibliography}
\end{document}